\documentclass[
aps,
nofootinbib,
prd,
superscriptaddress,
tightenlines,
notitlepage,
twocolumn,
showpacs,
floatfix
]{revtex4-1}
\usepackage{amsmath}
\usepackage{latexsym}
\usepackage{amsfonts}
\usepackage{graphicx}
\usepackage{mathrsfs}
\usepackage{epstopdf}
\usepackage{subfigure}
\usepackage{hhline}
\usepackage[utf8]{inputenc}
\usepackage{CJK}
\usepackage{float}
\usepackage{color}
\usepackage{hyperref}
\usepackage{diagbox}
\usepackage{xcolor}
\usepackage{bm}
\usepackage{lipsum}
\usepackage[normalem]{ulem}
\hypersetup{
    colorlinks=true,
    linkcolor=red,
    citecolor=blue,
}

\begin{document}

\title{Probing nontensorial gravitational waves with a next-generation
ground-based detector network}

\author{Jierui Hu }
\affiliation{School of Physics, Peking University, Beijing 100871, China}

\author{Dicong Liang}
\email[Corresponding author: ]{dcliang@pku.edu.cn}
\affiliation{Kavli Institute for Astronomy and Astrophysics, Peking University,
Beijing 100871, China}

\author{Lijing Shao}
\email[Corresponding author: ]{lshao@pku.edu.cn}
\affiliation{Kavli Institute for Astronomy and Astrophysics, Peking University,
Beijing 100871, China}
\affiliation{National Astronomical Observatories, Chinese Academy of Sciences,
Beijing 100012, China}

\date{\today}

\begin{abstract}
In General Relativity, there are only two polarizations for gravitational waves.
However, up to six polarizations are possible in a generic metric theory of
gravity. Therefore, measuring the polarization content of gravitational waves
provides an efficient way to test theories of gravity.  We analyze the
sensitivity of a next-generation ground-based detector network to nontensorial
polarizations.  We present our method to localize GW signals in the
time-frequency domain and construct the model-independent null stream for events with known sky
locations. We obtain results based on simulations of binary neutron star mergers
in a six-detector network.  For a single event at a luminosity distance
$D_L=100 \, {\rm Mpc}$, at $5\sigma$ confidence, the smallest amplitude for detection of scalar and
vector modes relative to tensor modes are respectively $A_{s}=0.045 $ and
$A_{v}=0.014 $. For multiple events in an averaged observing run of 10 years, the
detection limits at $5\sigma$ confidence are $A_s=0.05$ and $A_v=0.02$.  If we are fortunate, a few
strong events might significantly improve the limits. 
\end{abstract}

\maketitle

\section{Introduction}

Since the first detection of gravitational waves (GWs) in 2015
\cite{LIGOScientific:2016aoc}, various tests of gravity are implemented by
studying the data from ground-based detectors \cite{LIGOScientific:2016lio,
LIGOScientific:2018dkp, LIGOScientific:2019fpa, LIGOScientific:2020tif,
LIGOScientific:2021sio, Yunes:2016jcc, Berti:2018cxi, Berti:2018vdi,
Sathyaprakash:2019yqt, Kalogera:2021bya}.  One of the most important tests is
the polarization test. In General Relativity (GR), there are only two tensor
modes, namely the plus mode and the cross mode.  However, in a generic metric
theory of gravity, up to six polarization modes are permitted
\cite{Eardley:1973zuo, Eardley:1973br}, including two tensor modes, two vector
modes (vec-x mode and vec-y mode), and two scalar modes (breathing mode and
longitudinal mode).  In the Brans-Dicke theory, there is only one extra mode in
addition to the tensor modes, which is the breathing mode \cite{Eardley:1973zuo,
Eardley:1973br}.  While in the $f(R)$ theory and more general scalar-tensor
theories, the scalar mode presents in the form of a mixture of the breathing
mode and longitudinal mode \cite{Hou:2017bqj, Liang:2017ahj}. The two vector
modes appear in many vector-tensor theories \cite{Jacobson:2004ts, Gong:2018cgj,
Dong:2023xyb, Liang:2022hxd} and the polarization content can be anisotropic
when there is a nonzero spatial component in a Lorentz-violating vector
background \cite{ Kostelecky:2003fs, Liang:2022hxd,
Bailey:2023lzy}.  Considering these varieties, searching for extra polarizations
that differ from GR in GW signals can be a useful tool to test theories of
gravity. 

Different methods to search for the existence of extra polarizations in GWs were
developed in the literature.  Model selection based on Bayesian inference has
been applied to compact binary coalescences (CBCs) \cite{Isi:2017fbj,
LIGOScientific:2017ycc}, continuous waves (CWs) \cite{Isi:2017equ} and also GW
background (GWB) \cite{Callister:2017ocg, LIGOScientific:2018czr,KAGRA:2021kbb}. 
{Using the Fisher matrix method, the separability of the polarizations was studied in Refs.~\cite{Takeda:2018uai,Takeda:2019gwk}.
A morphology-independent test of extra polarizations was proposed by~\citet{Chatziioannou:2021mij}, which uses \texttt{BAYESWAVE} \cite{Cornish:2014kda,Cornish:2020dwh} to model different polarizations with a sum of sine-Gaussian wavelets. } 
Another straightforward method is to remove the tensor polarizations from the data and then
check if extra polarizations exist.  Null stream, which uses a linear
combination of outputs of multiple detectors to completely eliminate the
tensorial GW signals, was proposed to discriminate GW signals from noise
glitches in the context of GR \cite{Guersel:1989th, Chatterji:2006nh}.  This
idea was later extended to the null tests of GR \cite{Chatziioannou:2012rf,
Hayama:2012au} and recently was further developed in
Refs.~\cite{Hagihara:2019ihn, Pang:2020pf, Wong:2021cmp, LIGOScientific:2020tif,
LIGOScientific:2021sio, Zhang:2021fha}.  The null stream can be constructed only
for one single GW source, while for GWB produced by many sources, one has to
make use of the correlation between detectors.  Thus, the elimination method is
different and the details can be found in Refs.~\cite{Nishizawa:2009bf,
Nishizawa:2009jha, Omiya:2020fvw, Amalberti:2021kzh}.

In this paper,  we extend the analysis by \citet{Pang:2020pf} to the
next-generation ground-based GW detector network, including  Einstein Telescope
(ET),  Cosmic Explorer (CE), and so on.   ET is proposed in Europe, which
consists of three co-located detectors with 10\,km arm length in a triangular
geometry \cite{Punturo:2010zz}.   CE is proposed in America with 40\,km arm
length in L-shape \cite{Reitze:2019iox}, {and a 20 km CE-like detector is proposed in Australia \cite{Evans:2021gyd}}.  The sensitivity of the next-generation
detectors, like ET and CE, would have a factor of ten improvement over that of
the second-generation detectors \cite{LIGOScientific:2016wof}, which makes them
much easier to identify GW signals from the noise background.  We develop a new
method in Sec.~\ref{Select pixel points} to locate the GW signals in the
time-frequency domain more accurately.  Assuming that tensor modes always
present in GW signals, we construct null stream to remove the signals parallel
to tensor modes, in order to analyze the extra polarization components. {No specific waveform models are required when constructing the null stream.} Using
the statistical property of the energy of the null stream, we quantify the
possibility of the tensor-only hypothesis and estimate the sensitivity of the
next-generation GW detector network to extra polarizations. 

The paper is structured as follows.  We present the method to get the null
energy \cite{Pang:2020pf} in Sec.~\ref{section Null Energy}, and show our method
to locate GW signals in time-frequency domain in Sec.~\ref{Select pixel points}.
Next, we introduce the parameterized form of possible extra polarizations in
Sec.~\ref{Injection of extra polarizations}.  Section~\ref{Network sensitivity}
defines the network sensitivity and presents a possible network configuration
used in our analysis.  The simulations of single-event test and multiple-event
test are performed in Sec.~\ref{Results and Discussions}, where we quantify the
sensitivity to extra polarizations using the relative amplitude.  We discuss the
results of the relative amplitude sensitivity and compare our multiple-event results
with those from the second-generation detector network obtained by
\citet{Pang:2020pf}.  Conclusions are drawn in Sec.~\ref{Conclusions}.

\section{Methodology}
\label{Methodology}

In this section, we follow \citet{Pang:2020pf} to present the method to
construct the null stream and get the null energy of a GW event with known
direction.  By using the statistical distribution of the null energy, we assign
a $p$-value to test the extra polarizations in the GW data.

In a generic metric theory of gravity, up to six polarizations are allowed for
GWs \cite{Eardley:1973zuo,Eardley:1973br}: two tensor modes ($+$ and $\times$),
two vector modes ($x$ and $y$) and two scalar modes ($b$ and $l$). Thus, the
strain of GWs recorded at an interferometric detector can be expressed as a
linear combination of these six modes \cite{Will:2014kxa},
\begin{equation}
    s_{\rm GW} (t) = \sum\limits_A F^A h^{{A}} (t),
\end{equation}
where $A \in \{ +,\times,x,y,b,l \}$ represents the six polarizations,  $h_A
(t)$ is the strain of polarization $A$ and $F^A$ is its beam pattern function,
\begin{equation}
    F^A = D^{ij} e_{ij}^A.
\end{equation}
In the above equation $e_{ij}^A$ is the polarization tensor and the detector
tensor $D^{ij}$ in the low frequency limit is given by
\cite{Nishizawa:2009bf,Liang:2019pry},
\begin{equation}
    D^{ij}= \frac{1}{2} (\hat{u}_i \hat{u}_j -\hat{v}_i \hat{v}_j),
\end{equation}
where $\hat{u}$ and $\hat{v}$ are the unit vectors of two arms of the detector. 
Due to the degeneracy between $F^b$ and $F^l$ in the low frequency limit, it is
unlikely to distinguish the two scalar modes for the ground-based detectors
\cite{Nishizawa:2009bf,Liang:2019pry}. Thus, we use breathing mode to represent
scalar modes in our analysis.

\subsection{Null energy}
\label{section Null Energy}

Let us consider a network of $D$ interferometric detectors labeled by $j=0,...,
D-1$. Here we assume that the GW signals have only tensor modes. The strain data
of GW detectors in the frequency domain can be expressed in the following matrix
form,
\begin{equation}
\begin{aligned}
    &\Tilde{\boldsymbol{d}}(f)=\Tilde{\boldsymbol{s}}(f)+\Tilde{\boldsymbol{n}}(f),
    \\
    &\Tilde{\boldsymbol{s}}(f)=\boldsymbol{F}\Tilde{\boldsymbol{h}}(f),
\end{aligned}\label{strain data}
\end{equation}
where
\begin{equation}
    \Tilde{\boldsymbol{d}}=\begin{pmatrix}\Tilde{d}_0 \\ \vdots \\
    \Tilde{d}_{D-1}\end{pmatrix}
    , \quad
    \Tilde{\boldsymbol{h}}=\begin{pmatrix}\Tilde{h}^{+}  \\
    \Tilde{h}^{\times}\end{pmatrix} , \quad
    \Tilde{\boldsymbol{n}}=\begin{pmatrix}\Tilde{n}_0 \\ \vdots \\
    \Tilde{n}_{D-1}\end{pmatrix},
\end{equation}
and
\begin{align}
    \boldsymbol{F} &= \begin{pmatrix}
        \boldsymbol{F}^{+} \,, ~ \boldsymbol{F}^{\times}
    \end{pmatrix}, \\
    \boldsymbol{F}^A &= \begin{pmatrix}
        F^{A}_0 \\
        \vdots \\
        F^{A}_{D-1}
    \end{pmatrix}.
\end{align}
Here, we use $F^A_{j}$ to denote detector $j$'s beam pattern function for
polarization $A$, and $\boldsymbol{F}\in \mathbb{R}^{D\times 2}$ is the beam
pattern matrix.  We further assume that the merger time of the binary in each
detector is shifted to be the same.

To facilitate the analysis of signal immersed in the noise, we whiten the data
as
\begin{equation}
    \Tilde{\boldsymbol{d}}_w(f)=\boldsymbol{F}_w(f)
    \Tilde{\boldsymbol{h}}(f)+\Tilde{\boldsymbol{n}}_w(f),
\end{equation}
where
\begin{equation}
\begin{aligned}
    &\Tilde{d}_{w,j}[k]=\frac{\Tilde{d}_j[k]}{\sqrt{{S_j[k]}/{2\Delta f}}},\\
    &F_{w,j}^A[k]=\frac{F_j^A}{\sqrt{{S_j[k]}/{2\Delta f}}},\\
    &\Tilde{n}_{w,j}[k]=\frac{\Tilde{n}_j[k]}{\sqrt{{S_j[k]}/{2\Delta f}}} .
\end{aligned}
\end{equation}
The subscript $w$ indicates the whitening procedure and $S_j[k]$ is the
one-sided power spectral density (PSD) of the noise in detector $j$ at frequency
bin $k$.

For GW sources whose exact sky positions are known from their electromagnetic
counterparts, we can project out the tensor components in the GW signal by
constructing a null projector for tensor modes,
\begin{equation}
    \boldsymbol{P}_{\rm null} {[k]}=\boldsymbol{I} -
    \boldsymbol{F}_w {[k]} (\boldsymbol{F}_w {[k]} ^{\dagger}
    \boldsymbol{F}_w {[k]} )  ^{-1}\boldsymbol{F}_w {[k]} ^{\dagger}\label{null
    projector},
\end{equation}
where $\boldsymbol{F}_w \in \mathbb{R}^{D\times 2}$ only contains beam pattern
functions for $+$ and $\times$ modes. Since $\boldsymbol{P}_{\rm null}$ can
project out the tensor signal in Eq.~(\ref{strain data}), the residue
$\Tilde{\boldsymbol{z}}$ after the projection can be written as
\begin{equation}
    \Tilde{\boldsymbol{z}}= \boldsymbol{P}_{\rm
    null}\Tilde{\boldsymbol{d}}_w=\boldsymbol{0}+\boldsymbol{P}_{\rm
    null}\Tilde{\boldsymbol{n}}_w
    \label{residue z}.
\end{equation}
Notice that, the above projection does not depend on the detailed GW waveform,
thus is valid for any tensorial signals. After the projection, the residue only
contains noise  and that is why it is called {\it null stream}. 

Next, we transform the residue $\Tilde{\boldsymbol{z}}(f)$ back to time domain
$\boldsymbol{z}(t)$ and then perform the normalized Wilson-Daubechies-Meyer
(WDM) transform \cite{Necula:2012zz} to get the energy distribution in the
time-frequency domain,
\begin{equation}
    \left\Vert \Tilde{\boldsymbol{z}}_{\tau,{k}} \right\Vert ^2=\frac{ \left\Vert
    \sum\limits_{{k'}\in \mathbb{Z}  } e^{i\pi {k'} {k}/M} \boldsymbol{z}[\tau M + {k'}]
    \phi[{k'}] \right\Vert ^2}{\sum\limits_{{k'} \in \mathbb{Z}  }   \left\Vert \phi[{k'}]
    \right\Vert ^2},
\end{equation}
where $\tau$ and $k$ are the time and frequency indices, $M$ is the number of
frequency bins and $\phi[k']$ is the window function in the WDM transform. Here
we apply the WDM transform because of its superior time-frequency localization
capability.  The transform is normalized such that each pixel in the
time-frequency domain of white noise follows a standard normal distribution. A
simple WDM transform written in python can be found
online.\footnote{\url{https://github.com/Ecthelion666/WDM-transform-in-python}}

In general, the signal energy in the gravitational waves from a compact binary
coalescence in a circular orbit is encoded in a narrow region in the
time-frequency domain.  The null projection can wipe out all the energy of the
tensorial signals, while it does not affect the statistical property of the
noise.  From this perspective, it is more sensitive to compare the statistics
before and after the null projection for those in the narrow region where the signals locate, instead of the entire region in the time-frequency
domain.  Fortunately, the next-generation GW detectors have much higher
sensitivity than their ancestor generations, which makes it easier to
distinguish GW signal pixels from the noise background.  It is crucial to apply
the process of localizing pixel points to fully utilize the high sensitivity of
the next-generation GW detectors.  Thus, we only consider the null energy,
\begin{equation}
E_{\rm null}=\sum\limits_{\tau,k \in S_{\rm GW}} \left\Vert
\Tilde{\boldsymbol{z}}_{\tau,k } \right\Vert ^2
\label{null energy},
\end{equation}
where $S_{\rm GW}$ is a set of time-frequency indices that represent the
location of the GW signal. If we get the true sky position and there are only
tensor modes in the GW signal, the null energy follows a $\chi^2$ distribution
with the number of degrees of freedom (DOFs), $n_{\rm DOF}=N_{\tau k}(D-2)$,
where $N_{\tau k}$ is the number of pixels selected in the time-frequency domain
\cite{Pang:2020pf}.

\subsection{Localizing signal pixels}
\label{Select pixel points}

For a single detector, the signal-to-noise ratio (SNR) of a signal
$\Tilde{s}(f)$ is defined as
\begin{equation}
    \rho = \sqrt{4 \int_0^\infty \frac{\Vert \Tilde{s}(f) \Vert^2}{S(f)} df} \label{single snr},
\end{equation}
where $\Tilde{s}(f)$ is the observed signal in the frequency domain and $S(f)$
is the one-sided  {noise} PSD. The network SNR in $N$ detectors is defined as
\begin{equation}
    \rho_{\rm net} =    \sqrt{\sum\limits_{j=1}^N \rho^2_j}\label{network snr} \ .
\end{equation}

When the network SNR is high enough (say, $\rho_{\rm net}\gtrsim 100$), the GW
signal in the time-frequency domain---after the normalized WDM transform---can
be easily identified from the noise background.  On one hand, the signal pixels
contain a larger power than most noise pixels.  On the other hand, the signal
pixels are next to each other in the time-frequency plane while the loud noise
pixels distribute randomly.  Therefore, we devise the following method to select
signal pixels in the noisy data.

\begin{figure*}
    \centering
    \includegraphics[scale=0.58]{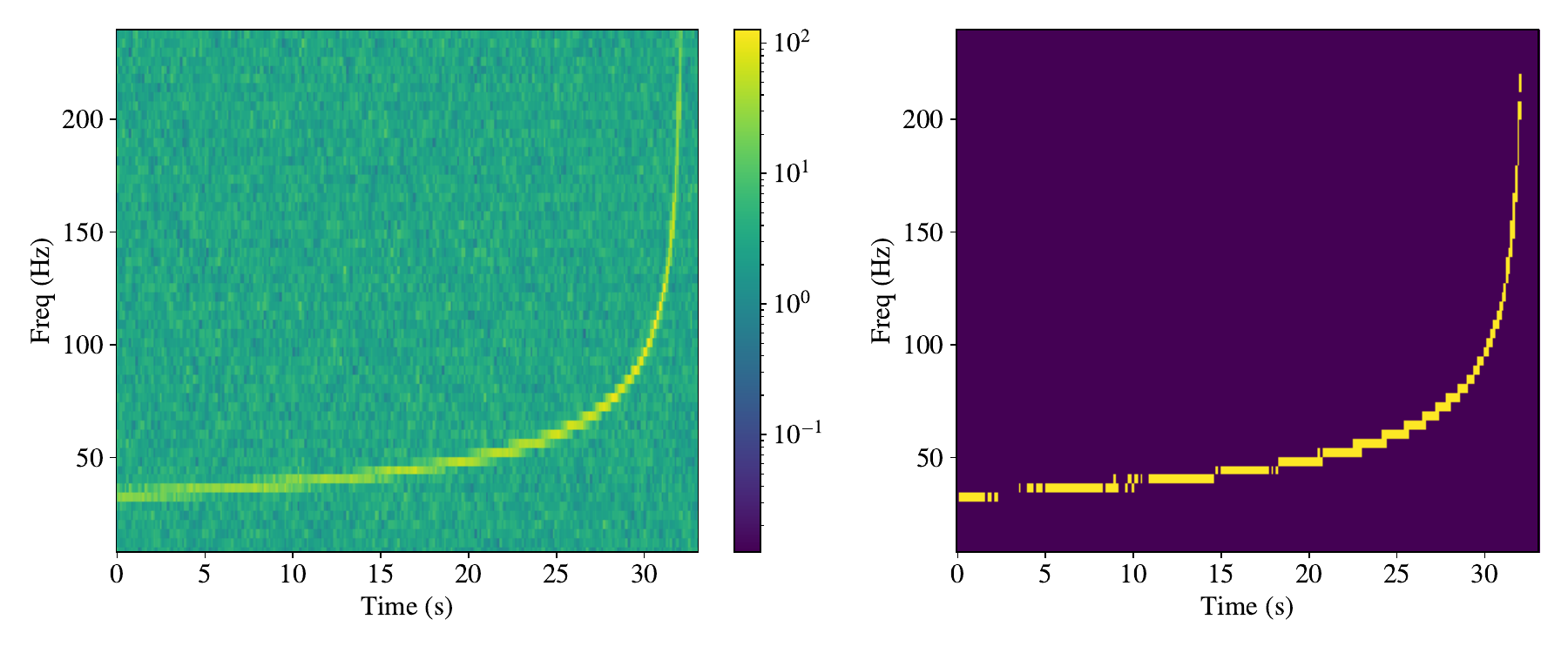}
    \caption{An example of localizing GW signal. The network SNR is $\rho_{\rm
    net}=141.2$ in this case. The left panel shows the result of the normalized
    WDM transform, while the right panel shows the pixels that are selected for
    the polarization test.}
    \label{fig:select pixel points}
\end{figure*}

We use $\rm IMRPhenomD\_NRTidalv2$ waveform model \cite{Dietrich:2019kaq} to
generate the tensorial signals.  The configuration of the detector network is
described in detail in Sec.~\ref{Network sensitivity}.  Figure~\ref{fig:select
pixel points} shows an example of localizing pixels for a GW signal from a
merger of non-spinning binary neutron star (BNS) with  masses $m_1=m_2=1.4
\,M_{\odot}$ at a luminosity distance of 450 Mpc. Its network SNR is $\rho_{\rm
net}=141.2$.  The method to select signal pixels is as follows. First, we
compute the root mean square normalized WDM transform result from the data of each detector
to get the combined WDM transform result, as shown in the left panel of
Fig.~\ref{fig:select pixel points}.  Second, to represent the GW signal, we
select a certain number of the brightest pixels and tag them ``good'' pixels. 
However, this procedure might include some loud noise pixels.  As an attempt to
remove those noise pixels, we only choose the pixels with five or more ``good''
pixels in the 7 neighbor pixels in the horizontal, vertical, and diagonal
directions because, as mentioned earlier, noise pixels are usually less
clustered than GW signals.  Finally, we obtain the time and frequency indices of
the pixels  where the GW signals locate, i.e. we obtain the set $S_{\rm GW}$ in
Eq.~(\ref{null energy}), as shown in the right panel of Fig.~\ref{fig:select
pixel points}.

Due to the high sensitivity of the next-generation detectors, we will observe
the early inspiral of BNS signals.  Compared to the late inspiral and the merger
stage, the early inspiral lasts for much longer time and the energy encoded in
this stage is relatively weaker.  Note that including weaker GW data in our
analysis may dilute the statistical significance of stronger GW data and is
hardly helpful to improve the sensitivity of the test.  In practice, we balance
the computational efficiency with the length of the data and optimize the number
of selected pixels for a better statistical significance.  From this
perspective, the signal we use in this paper is from the last 32 seconds before
merger plus 1 second after merger. Our sampling rate is $1024$ Hz.

\subsection{Injection of extra polarizations}
\label{Injection of extra polarizations}

Now, let us consider cases where extra polarizations are included in the GW
signal. The signal $\Tilde{\boldsymbol{s}}(f)$ in Eq.~(\ref{strain data})
changes,
\begin{equation}
    \Tilde{\boldsymbol{s}}(f)=\boldsymbol{F}^{+} \Tilde{h}^{+}(f)
    +\boldsymbol{F}^{\times}\Tilde{h}^{\times}(f)+\sum\limits_{e}
    \boldsymbol{F}^{e}\Tilde{h}^{e}(f),
\end{equation}
where the supscript $e$ denotes extra polarizations $\{x, y, b\}$. The null
projector in Eq.~(\ref{null projector}) can not project out extra polarizations
completely, as a consequence, the residue $\Tilde{\boldsymbol{z}}$ becomes
\begin{equation}
    \Tilde{\boldsymbol{z}}=\boldsymbol{P}_{\rm
    null}\Tilde{\boldsymbol{d}}_w=\boldsymbol{P}_{\rm
    null}\Tilde{\boldsymbol{n}}_w+\sum\limits_{e} \boldsymbol{P}_{\rm
    null}\boldsymbol{F}^{e}_w\Tilde{h}^{e}.
\end{equation}
In general, the introduction of extra polarizations will lead to an increase in
the expected value of the null energy, comparing to the noise-only data. 
Therefore, the null energy calculated by Eq.~(\ref{null energy}) no longer
follows the $\chi^2$ distribution,  and can be used as a statistic to quantify
the significance of extra polarizations. 

To test the sensitivity of next-generation GW detectors to extra polarizations,
we inject scalar and vector modes into the GW signal separately.  
For simplicity, we use $\boldsymbol{F}^{b} \in \mathbb{R}^{D\times 1}$ as the beam
pattern function for the scalar mode because of the degeneracy between the
response of breathing mode and longitudinal mode.  Furthermore, We assume that
the scalar modes are proportional to the plus mode with a relative amplitude
$A_s$, which can be interpreted as combining both the inclination dependence of
the scalar mode and the relative strength between the scalar component and the
tensorial component ${h}^{+}$ in a specific gravity theory
\cite{Takeda:2018uai}.  Hence, the scalar mode can be expressed as
\begin{equation}
    \Tilde{\boldsymbol{s}}_s(f) = \boldsymbol{F}^{b} A_s
    \Tilde{h}^{+}(f)\label{injection of scalar modes}.
\end{equation}
Similarly, we generate the two vector modes from the two tensor modes with a
relative amplitude $A_v$,
\begin{equation}
    \Tilde{\boldsymbol{s}}_v(f) = \boldsymbol{F}^{x}  A_v
    \Tilde{h}^{+}(f)+\boldsymbol{F}^{y} A_v
    \Tilde{h}^{\times}(f)\label{injection of vector modes}.
\end{equation}
Here we have assumed specific proportionalities for the $x$ and $y$ vector
modes. It is easy to release these assumptions. 

As we know, in some modified gravity theories, extra polarizations might have
different frequency evolution behaviours from the tensor modes, thus they
introduce multiple harmonics \cite{Chatziioannou:2012rf, Zhang:2019iim,
Liu:2020moh, Higashino:2022izi}.  To test the existence of such harmonics, we do
not need to apply the null projector to get the null energy because there is a
clear distinction in the location for different harmonics in the time-frequency
plane.  We can use the frequency evolution between different harmonics to
localize extra polarization components and then check if there is any
significant excess of energy.  The methodology is similar and we will not
discuss these cases here. In general, these are easier to search for as GW data
analysis is more sensitive to phase evolution.  When the extra polarizations
have the same frequency evolution as the tensor polarizations, they overlap with
each other in the time-frequency domain. Then we have to use the null projection
to remove the tensorial component. In this sense, the cases studied in this work
is conservative.

For GW events with known sky position, we can get the null energy using
Eq.~(\ref{null energy}) and assign a $p$-value to the tensor-only hypothesis,
\begin{equation}
    p=\int^{\infty}_{E_{\rm null}} \chi^2_{\rm DOF}(x) dx.\label{pvalue}
\end{equation}
Under this null hypothesis, $p$ is uniformly distributed between 0 and 1. After
injecting a non-zero extra polarization signal like in Eq.~(\ref{injection of
scalar modes}) or Eq.~(\ref{injection of vector modes}), the distribution of $p$
will deviate from the null hypothesis because the presence of extra
polarizations will enlarge the null energy.  Therefore, a small $p$-value
suggests a deviation from GR. In this work, we mainly use this $p$-value to
evaluate the sensitivity of a next-generation detector network to extra
polarizations.

\begin{table*}
    \centering
    \caption{Geometric angles for GW detectors in the detector network \cite{Gossan:2021eqe},  {including the latitude $\Lambda$,  longitude $\lambda$,
bisector angle $\gamma$ and opening angle $\zeta$ between detector's arms. The network comprises two CEs located at the same locations as LIGO Hanford and LIGO Livingston, one CE South located in the Southeastern Australia, and two pairs of ET-like detectors in a triangular configuration in Europe.}}
    \renewcommand{\arraystretch}{1.3}
    \begin{tabular}{p{5cm} p{3cm} p{3cm} p{3cm} p{2cm}}
        \hline\hline
        Facility &$\Lambda$ (deg)  & $\lambda$ (deg) &  $\gamma$ (deg)& $\zeta$ (deg)\\
        \hline
        CE (Hanford) & $46.455$ & $-119.408$ & $278.979$ & $90$\\
        CE (Livingston) & $30.563$ & $-90.774$ & $207.280$ & $90$\\
        CE {South} (Southeastern Australia) & $-35.000$ & $148.000$ & $272.450$ & $90$\\
        ET (Maastricht) & $50.754$ & $6.025$ & $247.761$ & $60$\\
        ET (Maastricht) & $50.754$ & $6.025$ & $127.761$ & $60$\\
        ET (Maastricht) & $50.754$ & $6.025$ & $7.761$ & $60$\\
        \hline
    \end{tabular}\label{configuration of the detector network}
\end{table*}

\subsection{GW detector network sensitivity}
\label{Network sensitivity}

Now we present the configuration of a next-generation GW detector network used
in this work. After that we introduce the optimal network sensitivity to extra
polarizations.

We define the optimal network sensitivity $\alpha^A(f,\hat{\Omega})$ before and after applying the null projection for polarization mode $A$ as
\begin{align}
    \alpha^A_{\rm before}(f,\hat{\Omega}) &=| \boldsymbol{F}_w^{A}(f) |, \\
    \alpha^A_{\rm after}(f,\hat{\Omega}) &=| \boldsymbol{P}_{\rm
    null}(f)\boldsymbol{F}_w^{A}(f) | .\label{sensitivity after projection}
\end{align}
Both $\alpha^A_{\rm before}(f,\hat{\Omega})$ and $\alpha^A_{\rm
after}(f,\hat{\Omega})$ are functions of the sky position $\hat{\Omega}$ because
$ \boldsymbol{P}_{\rm null}(f)$ and $\boldsymbol{F}_w^{A}(f)$ are functions of
$\hat{\Omega}$.  The optimal network sensitivities to scalar modes, vector modes
and tensor modes are defined as 
\begin{equation}
\begin{aligned}
    &\alpha^{\rm Scalar}(f,\hat{\Omega})=\alpha^b(f,\hat{\Omega}), \\
    &\alpha^{\rm Vector}(f,\hat{\Omega})=\sqrt{ \big[ \alpha^x(f,\hat{\Omega})
    \big]^2+ \big[\alpha^y(f,\hat{\Omega}) \big]^2}, \\
    &\alpha^{\rm Tensor}(f,\hat{\Omega})=\sqrt{ \big[\alpha^{+}(f,\hat{\Omega})
    \big]^2+ \big[\alpha^{\times}(f,\hat{\Omega})\big]^2}.
\end{aligned}
\label{ideal network sensitivity}
\end{equation}

To fully utilize the network of next generation detectors, in this paper, we consider a GW detector network, ``3ET+3CE'', consists of three
pairs of ET-like detectors located  at Maastricht in Netherlands \footnote{Here we treat ET as three independent detectors, for correlated noises, please refer to Ref.~\cite{Cireddu:2023ssf} }, and
three CE-like detectors located respectively at Hanford and Livingston in America, and Southeastern Australia. 
Table~\ref{configuration of the detector
network} gives the geometric factors (latitude $\Lambda$,  longitude $\lambda$,
bisector angle $\gamma$ and opening angle $\zeta$ between detector's arms) of
this detector network \cite{Gossan:2021eqe}. The azimuths are defined as clockwise relative to the due
North.

\begin{figure*}[!ht]
    \centering
    \includegraphics[width=\linewidth]{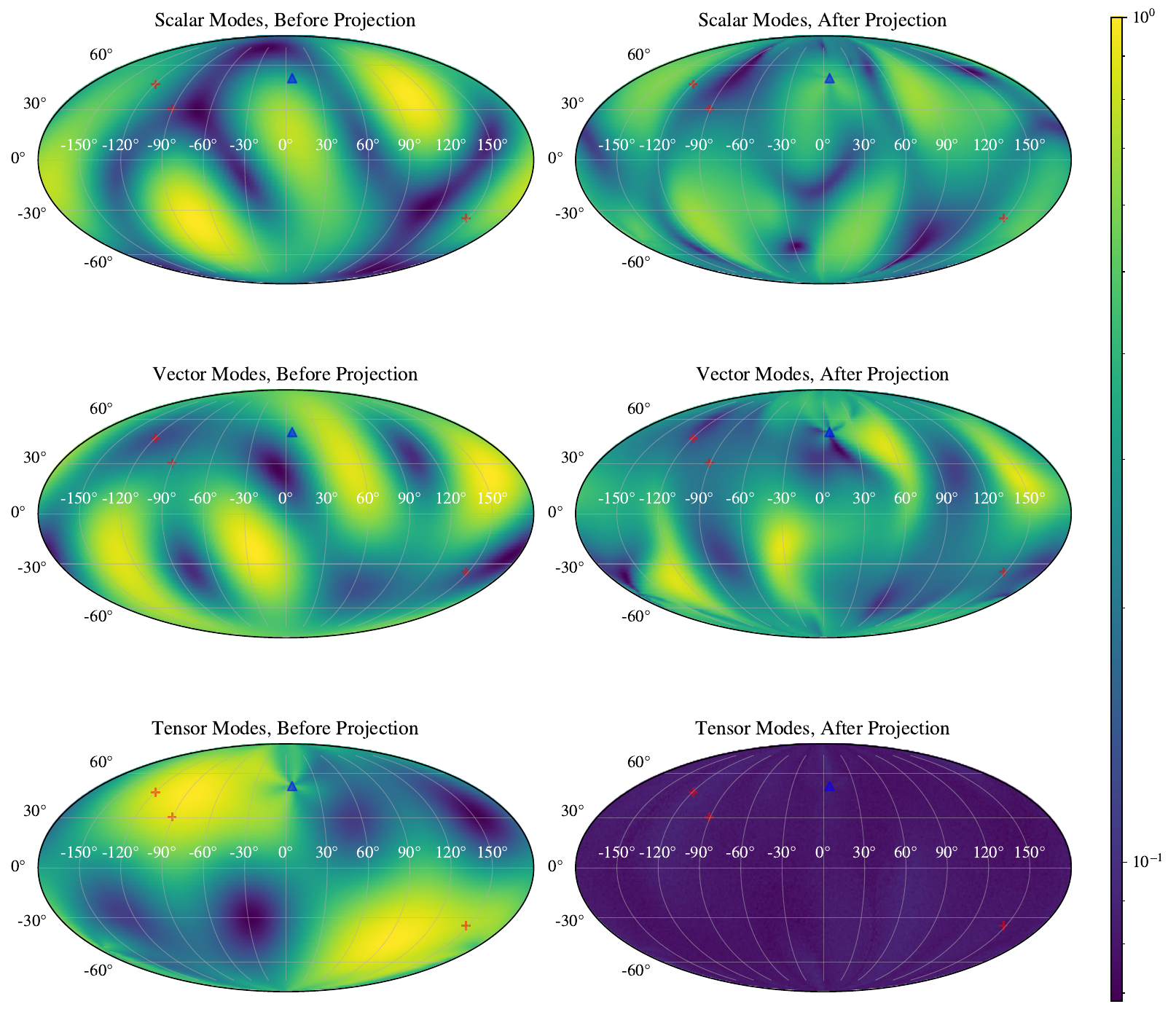}
    \caption{Normalized sky maps of optimal network sensitivity at 100\,Hz. The
    blue $\bigtriangleup $ is the location of the pair of ET-like detectors, and
    the red $+$ signs stand for three CE-like detectors. The plots in each
    polarization category are normalized by the maximum value of the optimal
    network sensitivity before projection, which means that the maximum values
    of the left panels are 1. The normalization factors are given in
    Eqs.~(\ref{eq:s:max}--\ref{eq:t:max}).}
    \label{fig:NetworkSensitivity}
\end{figure*}

For the sensitivities of CE-like and ET-like detectors, we use two analytical
PSDs, $\rm CosmicExplorerP1600143$ and $\rm EinsteinTelescopeP1600143$, in the
PyCBC package~\cite{Usman:2015kfa}. Figure~\ref{fig:NetworkSensitivity} shows
the normalized optimal network sensitivity sky maps of the ``3ET+3CE'' network
at 100\,Hz.  For each polarization mode, we normalize the sensitivity values
such that the maximum value of the optimal sensitivity before projection is
unity. The normalization factors are,
\begin{align}
    \alpha^{\rm Scalar}_{\rm before,max} &= 1.84\times10^{24} \,, \label{eq:s:max} \\
    \alpha^{\rm Vector}_{\rm before,max} &= 3.67\times10^{24} \,, \\
    \alpha^{\rm Tensor}_{\rm before,max} &=3.51\times 10^{24} \,. \label{eq:t:max}
\end{align}
From Fig.~\ref{fig:NetworkSensitivity}, we can see that the tensor modes are
completely projected out as expected and the sensitivity to the extra modes is
also deteriorated.  Besides, there are more residuals in vector modes after
projection than in scalar modes because the detectors' response of the vector
modes has less overlap with that of the tensor modes.

As we will see in the next section, $\alpha^A_{\rm after}(f,\hat{\Omega})$ can
serve as a computationally efficient indicator of the detectors' sensitivity to
extra polarizations and can be used as a guidance for optimizing future detector
network configuration for exra polarization searches.

\section{Results and Discussions}
\label{Results and Discussions}

With the methodology presented above, we give our results for single events in
Sec.~\ref{subsection Single event} and for multiple events in
Sec.~\ref{subsection Multiple events}.

\subsection{Single event}
\label{subsection Single event}

If the signal of extra polarization is very strong and there remains enough
energy after the null projection, then we can get an extremely small $p$-value
in a loud GW event. By setting the threshold $p$-value to {$p=2.9\times10^{-7}$ (the $5\sigma$ confidence)}, we want
to estimate the minimum relative amplitude $A_s$ or $A_v$---that we call {\it
relative amplitude sensitivity}---in different sky position.  When the actual
relative amplitude of the additional polarization components in the GW signal is
greater than the relative amplitude sensitivity, we are likely to confirm the
existence of extra polarizations in a single event.  Here we consider a BNS with
masses $m_1=m_2=1.4 \, M_{ \odot }$ and at a luminosity distance $D_L=100 \,
{\rm Mpc}$ for this study.

\begin{figure*}
    \centering
    \includegraphics[scale=0.58]{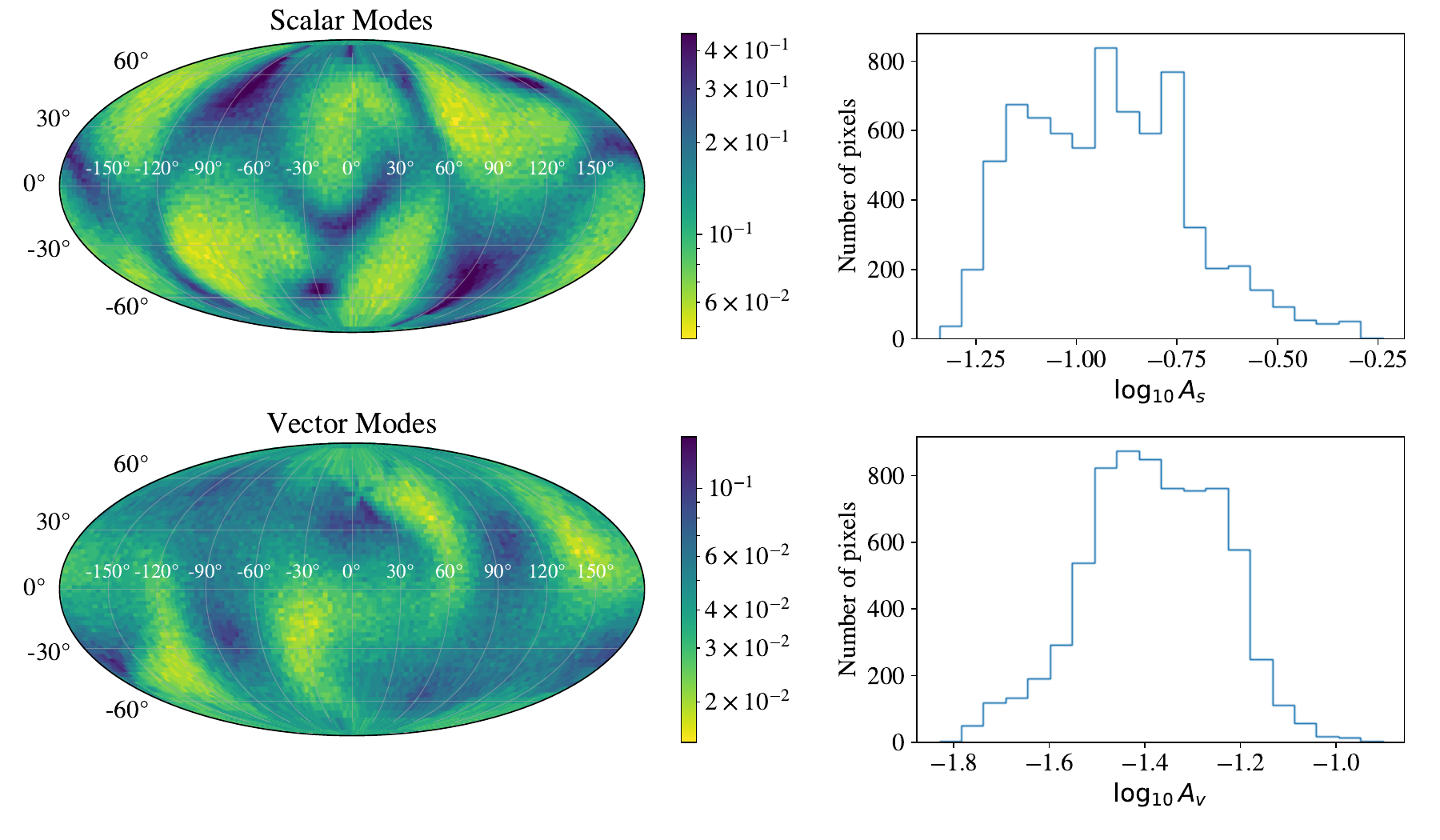}
    \caption{The left panels show the relative amplitude sensitivity sky maps
    for scalar modes and vector modes, where the relative amplitude sensitivity
    is defined as the minimum $A_s$ or $A_v$ in sky position $\hat{\Omega}$ for $p$-value to reach the $5\sigma$ threshold. The resolution of the event's sky location is 120
    pixels in longitude and 60 pixels in latitude, and each pixel has a same
    solid angle. The histograms on the right are the distribution of the minimum
    relative amplitudes. The minimum relative amplitude sensitivities for scalar
    modes and vector modes are given in Eqs.~(\ref{eq:min:s}--\ref{eq:min:v})
    and the mean values are given in Eqs.~(\ref{eq:mean:s}--\ref{eq:mean:v}).}
    \label{fig:askymap}
\end{figure*}

The left panels in Fig.~\ref{fig:askymap} show the relative amplitude
sensitivity sky map of the ``3ET+3CE'' network, and the right panels show the
distribution of the relative amplitude sensitivity.  The minimum relative
amplitude sensitivities for scalar modes and vector modes are {(at $5\sigma$ level)},
\begin{align}
    A_{s,\rm min} &=0.045\,,
    \label{eq:min:s} \\
    A_{v,\rm min} &=0.014\,.
    \label{eq:min:v}
\end{align}
The mean values are,
\begin{align}
    A_{s,\rm mean} &=0.133 \,, 
    \label{eq:mean:s} \\
    A_{v,\rm mean} &=0.043 \, .
    \label{eq:mean:v}
\end{align}
The relative amplitudes of vector modes are smaller than scalar modes for two
reasons. First, there is one more component in the vector modes [see
Eq.~(\ref{injection of vector modes})] than scalar modes [see
Eq.~(\ref{injection of scalar modes})].  Second, the vector modes have more
residuals than the scalar mode after applying the null projection (see the right
panels in Fig. \ref{fig:NetworkSensitivity}).

We can see that the pattern of scalar and vector modes in the sky maps of the
right panels in Fig.~\ref{fig:NetworkSensitivity} is similar to the sky maps in
Fig.~\ref{fig:askymap}.  This is because the null energy is proportional to the
square of ${\alpha^A_{\rm after}}(f,\hat{\Omega})$ and the $p$-value is
negatively correlated to the null energy.  Hence, the optimal sensitivity
${\alpha^A_{\rm after}}(f,\hat{\Omega})$ can be used to roughly estimate the
relative amplitude sensitivity, and these two quantities are negatively
correlated.  It is worth noting that the choice of frequency in ${\alpha^A_{\rm
after}}(f,\hat{\Omega})$ does not have a large impact on the result because the
PSD of the six detectors has similar shape in the sensitive frequency band.

\subsection{Multiple events}
\label{subsection Multiple events}

If the result in a single event is not significant enough to decide the
existence of extra polarizations, we can accumulate multiple GW events to
perform a combined analysis.  In this subsection we present simulations of
multiple events, which give a combined result from a long observing period.

As shown in Appendix \ref{Appendix A }, if the $p$-value of a single event $i$,
$p_i$, is uniformly distributed between 0 and 1, the statistic $S$ given by
\begin{equation}
    S=-2\sum\limits^{N}\limits_{i=1} \log p_i \label{statistic S}
\end{equation}
follows a $\chi^2$ distribution with $n_{\rm DOF} = 2N$. Therefore, the combined
$p$-value, $p_{\rm com}$, can be defined as
\begin{equation}
     p_{\rm com}=\int^{\infty}_{S} \chi^2_{2N}(x) dx\label{ combined p value},
\end{equation}
which characterizes the sensitivity of multiple events to extra polarizations.

In this multiple-event study, we generate GW events of BNSs randomly distributed
within a sphere that has a large radius with a luminosity distance $D_L\leq 1
{\rm Gpc}$. The masses of the neutron stars are simply assumed to be $1.4 \, M_{
\odot }$.  In reality, the mass of neutron stars has a relatively narrow
distribution compared to black holes, and it does not change significantly the
SNR of signals. As we can see in Figs.~\ref{fig:NetworkSensitivity} and
\ref{fig:askymap}, the sky location affects the sensitivity of the test
significantly, and the difference can be as large as one order of magnitude.
Therefore, our treatment to BNS masses suffice for this study and can be
extended in more dedicated studies in the future.  In our test, we only select
sources whose  $\rho_{\rm net}>350$ (within a radius $D_L \lesssim 200 \, {\rm
Mpc}$ approximately) to calculate $p_{\rm com}$ and these sources account for
$\sim 1\%$ of the total sources in our simulations.  The events with lower SNRs
are discarded from this analysis.

\begin{figure*}
    \centering
    \includegraphics[scale=0.5]{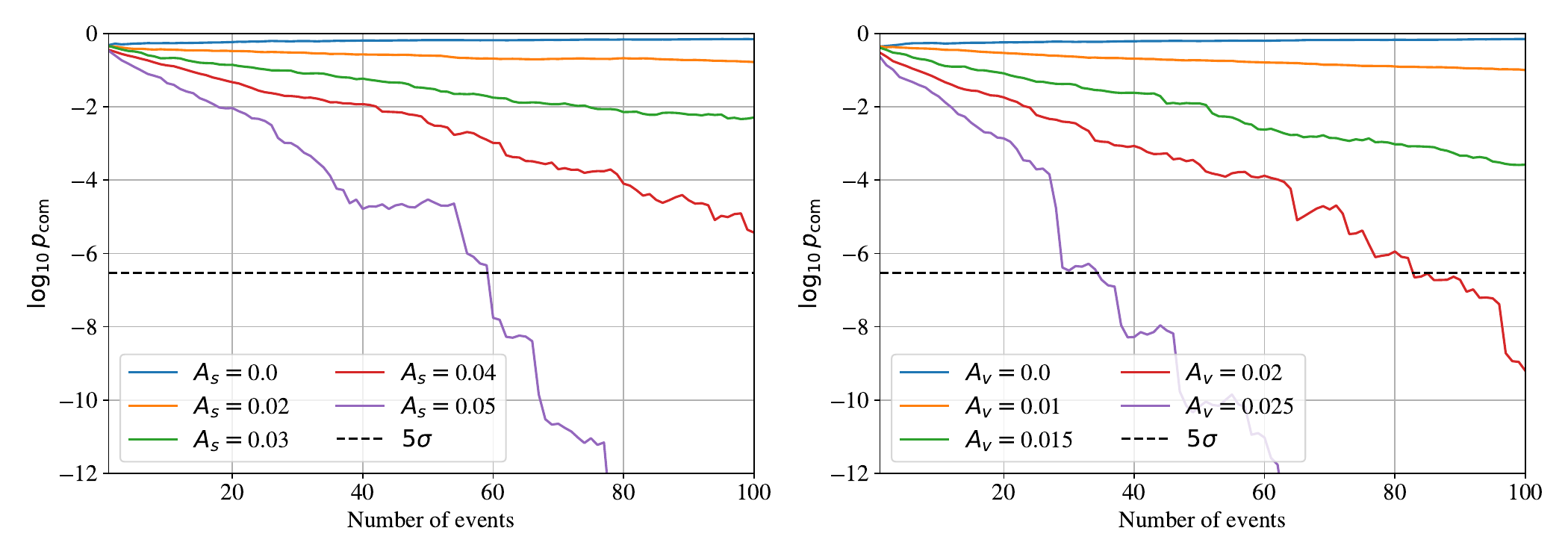}
    \caption{Combined $p$-value against the number of BNS events with $\rho_{\rm
    net}>350$. Each line denotes a different relative amplitude for scalar modes
    (left) or vector modes (right). The horizontal dashed line corresponds to the $5\sigma$ level. }
    \label{fig:pplot}
\end{figure*}

Figure \ref{fig:pplot} shows the quantity $\log_{10} p_{\rm com} $ against the
number of combined events, averaged over 100 simulations.  In our simulation,
when the merger rate of BNSs is chosen to be $1000 \, \rm Gpc^{-3} \, yr^{-1}$,
it will take about 10 years to generate about 100 candidate events with
$\rho_{\rm net}>350$.  With such an event rate, it would take about 6 years for
$A_s=0.05$ and 8 years for $A_v=0.02$ to reach the $5\sigma$ confidence level.  When
the relative amplitude gets smaller, it would take more time for $p_{\rm com}$
to hit the $5\sigma$ line.  The most recent inferred BNS merger rate is between
10 and 1700 $\rm Gpc^{-3} \, yr^{-1}$ \cite{KAGRA:2021duu}. Since the GW sources
that we analyzed are close enough, we can simply rescale the time scale
according to the realistic merger rate. 

We note that in Fig.~\ref{fig:pplot} there are some steep declines in the brown
lines in both panels ($A_s = 0.05$ for the scalar mode and $A_v = 0.025$ for the
vector modes). This is attributed to some rare events with small distances at
specific sky locations. As we can see in Sec.~\ref{subsection Single event}, the
$p$-value of a single event at $D_L=100 \, {\rm Mpc}$  might reach {the $5\sigma$ confidence}
when the relative amplitude is $A_{s,\rm min}=0.045$ or $A_{v,\rm min}=0.014$.
In practice, the significant contribution to the $p$-value from a single strong
signal may outweigh dozens of weaker signals. 

For the multiple-event test by \citet{Pang:2020pf}, the combined $p$-value of
the null hypothesis declines monotonically with the increasing numbers of GW
events.  It suggests that when the detection number is large enough, we cannot
tell apart the reason for an extremely small  $p_{\rm com}$ value, either due to
extra polarizations or the defects of the null-stream method.  While in our
simulations, the $p_{\rm com}$ value does not decrease monotonically, but
instead it is relatively stable after accumulating a large number of events.   
This is attributed to the high sensitivity of the next-generation detectors,
which allow us to locate the GW signals more precisely while avoiding mistakenly
selecting the noise pixels.

Notice that, the above $p$-value analysis can only indicate the existence of
extra polarizations, but cannot tell which extra polarizations are present in
the signals. Theoretically, it is possible to solve the equations,
$\Tilde{\boldsymbol{d}}(f) = \boldsymbol{F} \Tilde{\boldsymbol{h}}(f) +
\Tilde{\boldsymbol{n}}(f)$, to test extra polarizations when we have five
non-degenerate detectors. This is because that, due to the degenerate response
of GW detectors to the breathing mode and the longitudinal mode, there are only
five independent responses to all possible polarizations.  Such a study is out
of the scope of this work and we leave it to the future. 

\section{Conclusions}
\label{Conclusions}

In this work, we utilize the null stream to test the detectability of extra
polarizations in GWs with a next-generation detector network.  We introduce an
approach to improve the sensitivity of the test by locating GW signals in the
time-frequency domain precisely, though the exact sky locations of the GW
signals are needed to do so.  Using the high sensitivity of the next-generation
GW detectors, we can circumvent the influence of loud noise pixels to get a more
robust result, by looking for possible deviations via the $p$-value for the null
hypothesis that there are only two tensorial modes. It {extends} the method
presented  by \citet{Pang:2020pf}.

We use a detector network consisting of three  ET-like detectors and three CE-like
detectors for illustrative purposes.  We estimate the sensitivity of this
network for GW events coming from different sky positions by computing the
network sensitivity in Eq.~(\ref{sensitivity after projection}). It also
provides guidance for optimizing the geometry of GW detectors for extra
polarizations.  We simulate signals with extra polarizations and get the
detection thresholds for relative amplitudes, which are $A_{s,\rm min}=0.045$
for the scalar mode and $A_{v,\rm min}=0.014$ for vector modes, for a single BNS
event at a luminosity distance $D_L=100 \,{\rm Mpc}$ in an optimal sky location.
Finally, we combined the results from multiple events. For a 10-year observation
with BNS events uniformly distributed in the local Universe with an event rate
of $1000 \, \rm yr^{-1} \, Gpc^{-3}$, the averaged results indicate that the
detection limits of relative amplitudes are $A_{s,\rm min} \simeq 0.05$ for the
scalar mode and $A_{v,\rm min}\simeq 0.02$ for the vector modes. However, if we
are fortunate enough to observe some golden BNS events in the observing run,
this limit might become much better. 
{Besides, to obtain the results above, we have used the assumption that the frequency evolution for extra modes is parallel to tensor modes, which ignores the possibility of altered frequency evolution and multiple harmonics in some modified gravity theories \cite{Chatziioannou:2012rf}. Therefore, the results are conservative in this sense.
For the future study, it will be beneficial to use some specific waveform for extra polarizations, instead of using the GR waveform.  }

{It was proposed by~\citet{Takeda:2019gwk} that, the next generation detector can observe the early inspiral of BNSs for hours to days, thus the beam pattern function varies with time due to the Earth's rotation. In this sense, one detector is sufficient to separate different polarizations since it can be considered as several virtual detectors.
While in our work, we only concentrate on the late inspiral and merger of BNSs, which lasts for a much shorter time but contributes more to the total SNR. 
That is why we consider a network to probe extra polarizations. 
It will be intriguing to use a single next generation detector to construct null stream for the early inspiral of BNSs in the future and similar study for space-based detectors can be found in Ref.~\cite{Zhang:2019iim}.
It seems that~\citet{Takeda:2019gwk} gave a smaller estimation for the relative amplitude of the scalar mode to be detected. While the estimation error they obtained from Fisher matrix corresponds to $1\sigma$ confidence level, which is more relaxed than our work.
From this perspective, we think our results are comparable and we leave a further qualitative study to the future. }

Note that the sky positions are assumed to be known in our analysis, but in
practice the uncertainty of the sky positions can make the $p$-value smaller
even for the null hypothesis. Nevertheless, we rely on future electromagnetic
instruments to precisely locate nearby BNS events~\cite{KAGRA:2013rdx}.

\begin{acknowledgments}
We thank the anonymous referee for the useful comments and thank Yacheng Kang for helpful discussions.  This work was supported by the Beijing Natural Science Foundation (1242018), the
National Natural Science Foundation of China (11975027, 11991053), the China
Postdoctoral Science Foundation (2021TQ0018), the National SKA Program of China
(2020SKA0120300),  the Max Planck Partner Group Program funded by the Max Planck
Society, and the High-Performance Computing Platform of Peking University. J.H.\
was supported by the National Training Program of Innovation for Undergraduates
at Peking University.
\end{acknowledgments}

\appendix

\section{The statistic $S$} 
\label{Appendix A }

To prove the statistic $S$ in Eq.~(\ref{statistic S}) follows a $\chi^2$
distribution with $n_{\rm DOF} = 2N$, it is sufficient to prove that each
$-2\log p_i$ term in Eq.~(\ref{statistic S}) follows a $\chi^2$ distribution
with 2 DOFs. It is known that each $p_i$ is uniformly distributed in $[0,1]$.

Let us define $Y$ as
\begin{equation}
    Y=-2\log p.
\end{equation}
The probability density of $Y$ is $D(Y)$, which satisfies
\begin{equation}
    D(Y) dY = D(p) dp \label{DY},
\end{equation}
and $D(p)=1$ is the probability density of $p$.  The solution to Eq.~(\ref{DY})
is 
\begin{equation}
    D(Y) = \frac{1}{2} {e^{-\frac{Y}{2}}},
\end{equation}
which is the probability density of a $\chi^2$ distribution with 2 DOFs. Because
each $p_i$ in Eq.~(\ref{statistic S}) is independent to each other, the sum of
$-2\log(p_i)$ follows a $\chi^2$ distribution with $2N$ DOFs.

\bibliographystyle{apsrev4-1}
\bibliography{Ref}

\end{document}